\documentclass[aps,prx,reprint,superscriptaddress]{revtex4-2}
\usepackage{graphicx}
\usepackage{amsmath}
\usepackage{amssymb}
\usepackage{natbib}
\usepackage{hyperref}

\usepackage[dvipsnames]{xcolor}

\begin{document}





\title{Relaxation timescales and electron-phonon coupling in optically-pumped YBa$_2$Cu$_3$O$_{6+x}$ revealed by time-resolved Raman scattering.}


\author{N. Pellatz}
\affiliation{Department of Physics, University of Colorado, Boulder, Colorado 80309, USA}
\affiliation{Center for Experiments on Quantum Materials, University of Colorado - Boulder, Boulder, Colorado, 80309, USA}

\author{S. Roy}
\affiliation{Department of Physics, University of Colorado, Boulder, Colorado 80309, USA}
\affiliation{Center for Experiments on Quantum Materials, University of Colorado - Boulder, Boulder, Colorado, 80309, USA}

\author{J-W. Lee}
\author{J. L. Schad}
\affiliation{Department of Materials Science and Engineering, University of Wisconsin-Madison, Madison, WI 53706, USA}

\author{H. Kandel}
\author{N. Arndt}
\affiliation{Department of Physics and Mathematics, University of Wisconsin-Parkside, Kenosha, WI 53140}

\author{C.B. Eom}
\affiliation{Department of Materials Science and Engineering, University of Wisconsin-Madison, Madison, WI 53706, USA}

\author{A.F. Kemper}
\affiliation{Department of Physics, North Carolina State University, Raleigh, North Carolina 27695, USA}

\author{D. Reznik}
\email[Corresponding Author: ]{dmitry.reznik@colorado.edu}
\affiliation{Department of Physics, University of Colorado, Boulder, Colorado 80309, USA}
\affiliation{Center for Experiments on Quantum Materials, University of Colorado - Boulder, Boulder, Colorado, 80309, USA}




\begin{abstract}
 
Time resolved measurements provide a new way to disentangle complex interactions in quantum materials due to their different timescales. We used pump-probe Raman scattering to investigate the apical oxygen vibration in  YBa$_2$Cu$_3$O$_{6.9}$ under nonequilibrium conditions. Time-dependence of the phonon population demonstrated strong electron-phonon coupling. Most importantly, the phonon shifts to a higher energy due to transient smearing of the Fermi surface in a remarkable agreement with theory. We also discuss new insights into photoinduced superconductivity reported at lower doping that follow from these results.

\end{abstract}




\maketitle



Driving quantum materials with electromagnetic fields can generate novel phases and states away from thermal equilibrium \cite{Wang453, RevModPhys.82.2731, tengdin2018critical, PhysRevB.83.125102, ehrke2011photoinduced}. Recently reported signatures of superconductivity at elevated temperatures in photoexcited copper oxides and intercalated fullerenes are particularly interesting but still enigmatic \cite{hu_optically_2014,mitrano_possible_2016,budden2020evidence,Liu_2020,PhysRevB.90.100503,cremin2019photoenhanced,takabayashi2016unconventional}. In most of these experiments the ultrafast laser pulse (pump) drives or photoexcites the system and another ultrafast pulse (probe) takes snapshots  of a specific property as a function of pump-probe time delay. Such experiments also elucidate energy flows between phonons, electrons, and magnons providing a way to determine the strength of different interactions. 

This work focuses on relaxation timescales and interactions between electrons and phonons in a prototypical copper oxide superconductor, YBa$_2$Cu$_3$O$_{6+x}$. 

Electron-phonon coupling in the copper oxides is still enigmatic. It allows electron-hole recombination  and electron scattering with the creation or annihilation of phonons; phonons in turn can decay into electron-hole pairs. As a result, spectroscopic quasiparticle peaks shift and/or broaden due to decreased lifetime. However,
electron-phonon scattering is not the only process that gives rise to these effects. Internally, phonon-phonon coupling (anharmonicity) also broadens the phonon peaks; similarly, electron-electron interactions can broaden the electronic quasiparticles, as does disorder \cite{PhysRevB.84.214516}. Thus, extracting just the
electron-phonon coupling strength from the linewidths of quasiparticle peaks is challenging.

\begin{figure*}
\centering
\includegraphics[trim={0cm 0cm 0cm 0cm},width=0.9\textwidth]{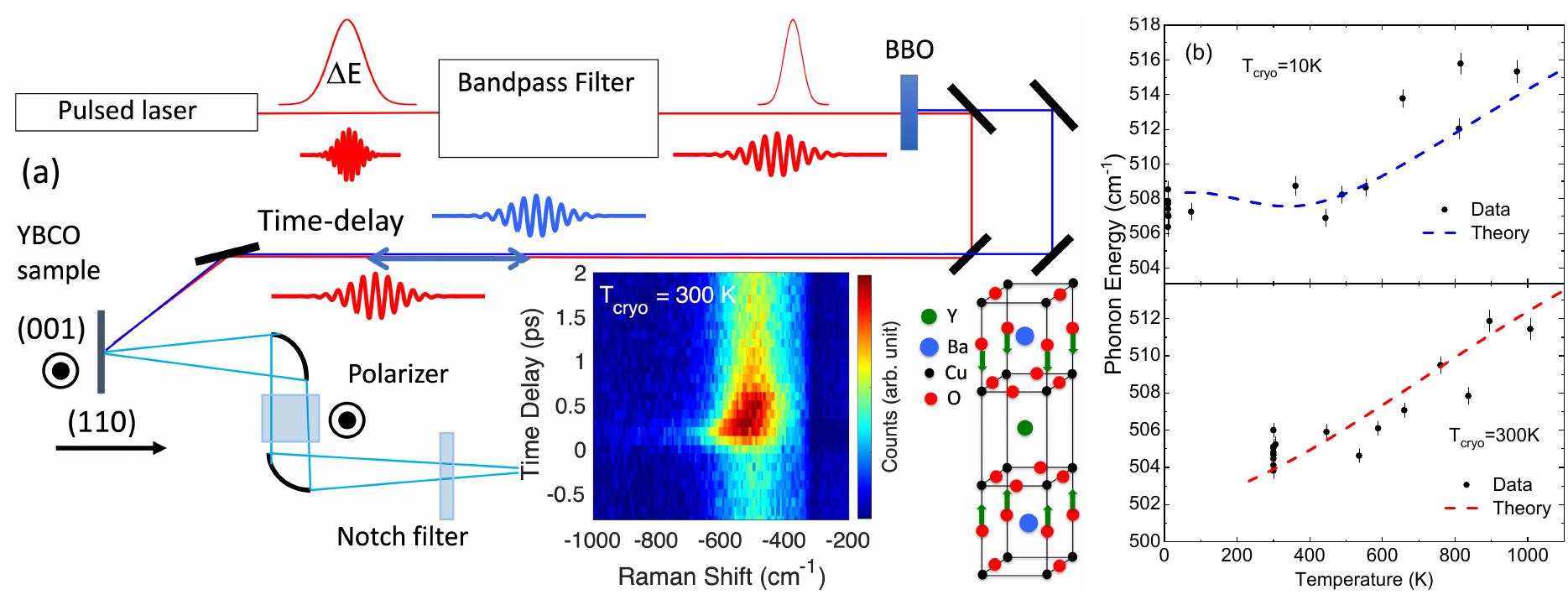} 
\caption{Time-resolved Raman scattering (TRR) setup and key results. (a) Schematic of the TRR experiment and the color map showing representative data on the anti-Stokes (AS) side of the spectrum obtained with the sample at 300 K. Negative time corresponds to the probe pulse arriving before the pump pulse when the system is still at thermal equilibrium.  Most intensity in the peak is from the apical oxygen phonon whose atomic displacements are indicated by green arrows in the schematic of the YBa$_2$Cu$_3$O$_{7}$ unit cell  to the right of the color map.  (b) Comparison of diagrammatic theory \cite{kemper2018general} and experiment for the dependence of phonon energy as a function of electronic temperature at delay times up to 600fs. The temperature in the cryostat was 10K/300K in the top/bottom panels respectively. Impurity scattering of 20meV was added to the theory curve in the upper panel, but not to the lower panel, since scattering by phonons at 300K is greater. Unrenormalized phonon energy was picked to obtain good agreement of theory with experiment at thermal equilibrium \cite{kemper2018general}. \label{schematics}}
\end{figure*}

To get around this problem, we used time-resolved Raman scattering (TRR) (Fig. \ref{schematics}a)\cite{Linde80,yang_novel_2017, PhysRevLett.54.2151, PhysRevB.80.121403, PhysRevLett.100.225503, kang2008optical, PhysRevB.81.165405, wu2012hot, PhysRevB.99.094305, Versteeg2018}. 
In TRR an ultrafast optical pump laser pulse first photoexcites the material and Raman scattering from another time-delayed pulse probes the system. In the copper oxides, previous TRR work highlighted ultrafast destruction of the antiferromagnetic order in the undoped parent compound of YBa$_2$Cu$_3$O$_{6+x}$ \cite{yang_ultrafast_2020} and nonequilibrium behavior of the superconducting gap in Bi$_2$Sr$_2$CaCu$_2$O$_{8+\delta}$ (BSCCO) \cite{PhysRevLett.102.177004}. 

In our experiment near infrared (IR) pump pulses create hot electrons, and the time-delayed Raman probe measures the apical oxygen $A_{1g}$ phonon. 
Apical oxygen modes recently came to the center of attention due to their impact on electronic  \cite{rosenstein2020apical}  and magnetic \cite{Peng2017} degrees of freedom of the copper-oxygen planes, interlayer charge transport \cite{michael2020parametric}, and photoinduced superconductivity \cite{hu_optically_2014,Liu_2020,PhysRevB.90.100503,cremin2019photoenhanced,rosenstein2020apical}. The $A_{1g}$ mode is known for a large spectroscopic linewidth \cite{limonov_raman_1998,altendorf_electron-phonon_1991,PhysRevB.43.13751} and an anomaly at the superconducting transition temperature, T$_c$. 

Upon optical pumping the phonon occupation number increased dramatically due to absorption of energy from photoexcited electrons as expected from strong electron-phonon coupling. \cite{Gunnarsson_2008,Johnston2010,ALEXANDROV2002650,Falter95,Falter93,Rosch2005}. This phonon also hardened at short time delays as a result of the decrease of its electronic self-energy expected from smearing of the Fermi surface due to very high transient electronic temperature. This effect provides a rigorous test of electron-phonon calculations based on electronic structure. Our results were in quantitative agreement with Green's functions-based theory of Refs. \cite{Johnston2010,kemper2018general} (see Fig. \ref{schematics}b). 



Data presented in this letter were collected on a 170 nm-thick (110)-oriented YBa$_2$Cu$_3$O$_{6.9}$ thin film prepared by pulsed laser deposition on a (110) LaAlO$_3$ substrate with T$_c$ of 81 K \cite{PhysRevB.44.12643,PhysRevB.52.6845}. We also obtained a dataset on a similar film with reduced oxygen concentration (T$_c$=40K) with the temperature in the cryostat of 100K. Our TRR setup (Fig. \ref{schematics}a) uses 20 kHz 790 nm (1.57 eV) laser pulses from an amplified mode-locked Ti:sapphire laser, which produces intense 40 fs pulses. Second harmonic generation at 395 nm (3.14 eV) was used as the probe light source for Raman scattering. 


The time-energy uncertainty principle limits the energy resolution of the ultrafast probe. In order to resolve the phonons from the elastic line, the 790 nm pulses were passed through an in-house built time-compensating band-pass filter (Fig. \ref{schematics}a) making both the pump and the probe pulses narrower in energy and broader in time. A cross-correlation measurement gave a time-resolution of 220 fs FWHM (see Supplementary Note 1 in \cite {Supp}). The time-averaged probe power was below 1 mW, to eliminate self-pumping nonlinearities (in Supplementary information \cite{Supp}, see reference \cite{PhysRevB.80.121403} therein).


The scattered light was collected by a pair of parabolic mirrors with a polarizer in the middle, and analyzed on a single-stage McPherson spectrometer equipped with a LN$_2$-cooled CCD detector. A custom-made notch filter blocked elastically-scattered light. The samples were in air or in a cryostat in a He exchange gas. We show results obtained with the pump photon polarization parallel to the ab-plane. Photon polarization along the c-axis gave similar results. Background measured under identical conditions but without the probe was subtracted from raw data. 




We read off phonon temperature from the intensities on the Stokes (S) and anti-Stokes (AS) sides, I$_S$ and I$_{AS}$ \cite{yang_novel_2017, PhysRevLett.54.2151}. They are related by the fundamental principle of detailed balance:
\begin{equation}
\frac{g\textrm{I}_{AS}}{\textrm{I}_{S}} = \frac{\left(\omega_0 + \omega_{ph}\right)^4}{\left(\omega_0 - \omega_{ph}\right)^4}\textrm{e}^{-\hbar\omega_{ph}/k_BT_{ph}},
\label{detailed-balance}
\end{equation}
where $\omega_0$ is the laser frequency, $\omega_{ph}$ is the phonon energy, $k_B$ is the Boltzmann’s constant, $T_{ph}$ is the phonon temperature, and $g$ is equal to 1. 
Introducing $g$ allows us to correct for the systematic error due to imprecise spectrometer calibration. We made $g = 0.9$ to make the phonon temperature at negative times equal to 300 K at room temperature. Following convention, we define the temperature of each bosonic mode, $T_{boson}$, via its relation to the occupation number $n = \left(\mathrm{e}^{\hbar\omega_{ph}/k_BT_{boson}}-1\right)^{-1}$. Away from thermal equilibrium different phonons have different occupation numbers, and therefore different temperatures. According to the fluctuation-dissipation theorem, the S and AS Raman intensities are given by $\textrm{I}_S = (n+1)\chi''\left(\hbar\omega\right)$, and $\textrm{I}_{AS} = n\chi''\left(\hbar\omega\right)$, where $\chi''$ is the imaginary part of the Raman response function (polarizability) of the phonon of interest. The color map in Fig. \ref{schematics}a showcases the dramatic increase of the AS intensity, indicating an increase of $T_{ph}$ (Eqs. 1,2), as well as a peak shift to larger energy  right after photoexcitation.

\begin{figure}
\centering
\includegraphics[trim={0cm 0cm 0cm 1cm},width=0.48\textwidth]{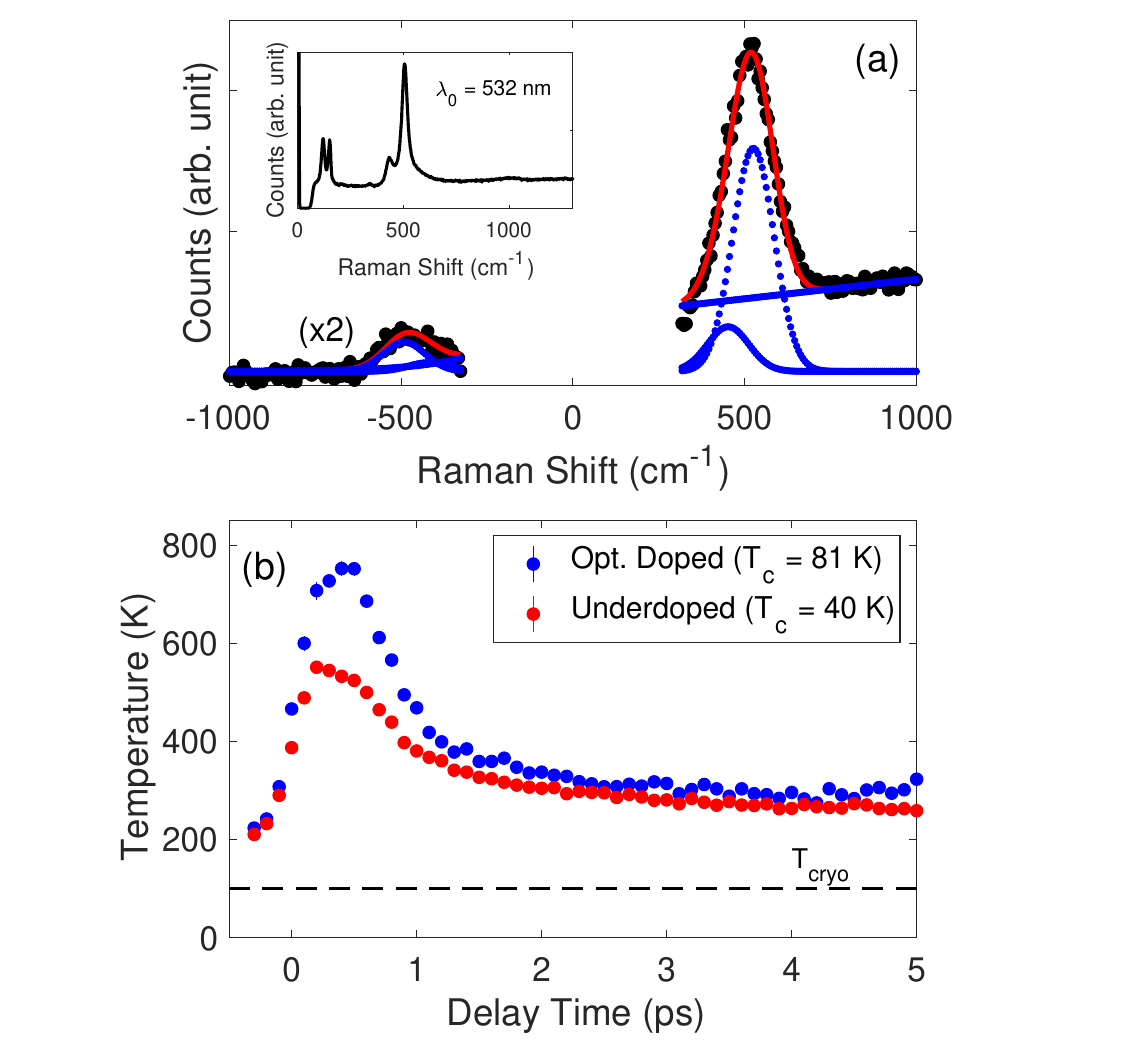} 
\caption{Raw data and phonon temperature at two doping levels. (a) Background-subtracted TRR spectra at different delay times with the cryostat at 300 K (black circles). The peaks at $\pm$500 cm$^{-1}$ are Stokes ($+$) and anti-Stokes ($-$) phonon peaks. Red line is a guide to the eye representing smoothed data without the pump. Large peak widths are due to the increased energy width of the pulsed laser. (b) Phonon temperature as a function of time after photoexcitation at optimal doping and reduced doping where photoinduced superconductivity was reported earlier \cite{Liu_2020}. Temperature in the cryostat was 100K. Pump fluence was 8mJ/cm$^2$, the same as in Ref. \cite{Liu_2020}. Note that pump energy is absorbed first by electron-hole pairs, which then thermalize with the apical oxygen phonons, i.e. electronic temperature never drops below the phonon temperatures. 
\label{raw}}
\end{figure} 

\begin{figure}
\centering
\includegraphics[trim={0cm 0cm 0cm 1cm},width=0.48\textwidth]{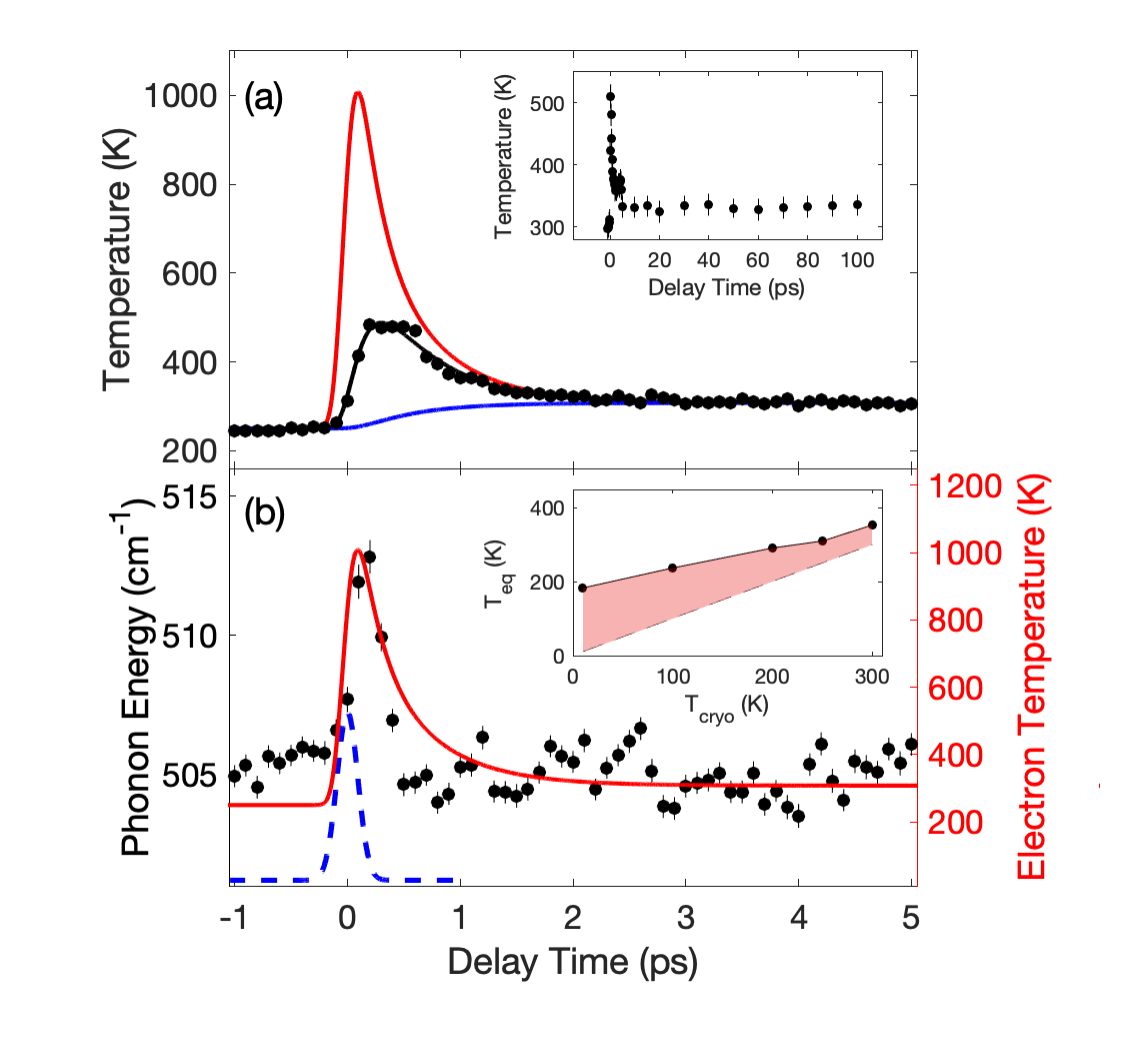} 
\caption{Phonon temperature and energy together with fits to the two temperature model. Pump (probe) fluence was 1.4 mJ/cm$^2$ (15 $\mu$J/ cm$^2$). $T_{cryo}$ was 250 K. Black line represents hot phonons, red -- hot electrons, blue -- cold phonons. The electronic temperature curves in (b) and (c) are the same.  Inset to (b) illustrates the behavior at large delay times for the sample in air at 300 K. Dashed blue line in (c) is the measured pump-probe cross correlation centered at t=0.
\label{temperatureandenergy}}
\end{figure}

The apical oxygen phonon around 500 cm$^{-1}$  and a weaker plane oxygen mode at 440 cm$^{-1}$ (Fig. \ref{raw}a inset), which may include substantial apical oxygen character \cite{PhysRevB.65.064501,MISOCHKO1990387}, dominate the Raman spectrum in the zz geometry where both incident and scattered photons are polarized perpendicular to the copper-oxygen planes (Fig. \ref{raw}a) \cite{limonov_raman_1998,altendorf_electron-phonon_1991,PhysRevB.46.11725}. The intrinsic lineshape of the phonon peaks is known to be best describes by characteristic Fano profiles (See inset in Fig. \ref{raw}a and \cite{limonov_raman_1998,altendorf_electron-phonon_1991,PhysRevB.46.11725}), however, due to a very broad energy resolution of the experiment, they were fit with Gaussians as shown in the main panel of Fig. \ref{raw}a. When fitting the time-resolved data where the peaks are not resolved (Fig. \ref{raw}a, main panel), the intensity of the 440 cm$^{-1}$ peak was fixed at 20\% of the 500 cm$^{-1}$ peak, consistent with Raman intensities obtained with a high energy resolution 360 nm laser whose wavelength is close to 395 nm of the probe pulsed laser. The linewidths of the two peaks are similar (Fig. \ref{raw}a inset), so we assumed similar electron-phonon coupling and kept the intensity ratio the same at all delay times. Assuming weak electron-phonon coupling for the 440 cm$^{-1}$ peak did not significantly change the fit results. 

Fig. \ref{raw}b shows the time-dependence of the phonon temperature for an optimally-doped and underdoped samples obtained under the pumping conditions used to generate transient superconductivity with the near IR pump in an earlier study \cite{Liu_2020}. The two samples behave similarly with smaller maximum temperature achieved by the underdoped sample. Note that in both cases the phonon temperature remains above 300K up to long delay times although the temperature in the cryostat was 100K.


We performed a detailed analysis of the time-dependent phonon data obtained with a lower pump fluence (Fig. \ref{temperatureandenergy}). The phonon temperature dramatically increases within the time-resolution to about 450 K independent of the temperature in the cryostat. This increase is followed by the exponential decay starting from 0.5 ps. The decay saturates around 5 ps (Fig. \ref{temperatureandenergy}a inset) when the photoexcited region reaches the internal thermal equilibrium temperature, $T_{eq}$ = 300 K. Inset to Fig. \ref{temperatureandenergy}b shows $T_{eq}$ as a function of $T_{cryo}$ highlighting increased transient heating of the sample, $\Delta T$, at lower temperatures due to its smaller heat capacity. Further equilibration with the cryostat exchange gas is much longer than 100 ps (see inset to Fig. \ref{temperatureandenergy}a). An earlier study attributed this heating to nonthermal effects, but the heating at these time delays is consistent with the pump fluence as discussed below.


Our 1.5 eV pump photons should create electron-hole pairs \cite{PhysRevB.47.8233}, which thermalize among themselves much faster than the time resolution. Time resolved angle-resolved photoemission (trARPES) showed  that electrons reached a maximum temperature of 800 K with 100 $\mu$J/cm$^2$ pump pulses in BSCCO \cite{PhysRevLett.99.197001}. In graphite the electronic temperature reached far above 1000 K when pumping with 150 $\mu$J/cm$^2$ \cite{yang_novel_2017, ishida2011non, PhysRevLett.102.086809}. It is reasonable to expect similarly high electronic temperatures in YBCO. 

\begin{figure}
\centering
\includegraphics[trim={0cm 0cm 0cm 0.5cm},width=0.45\textwidth]{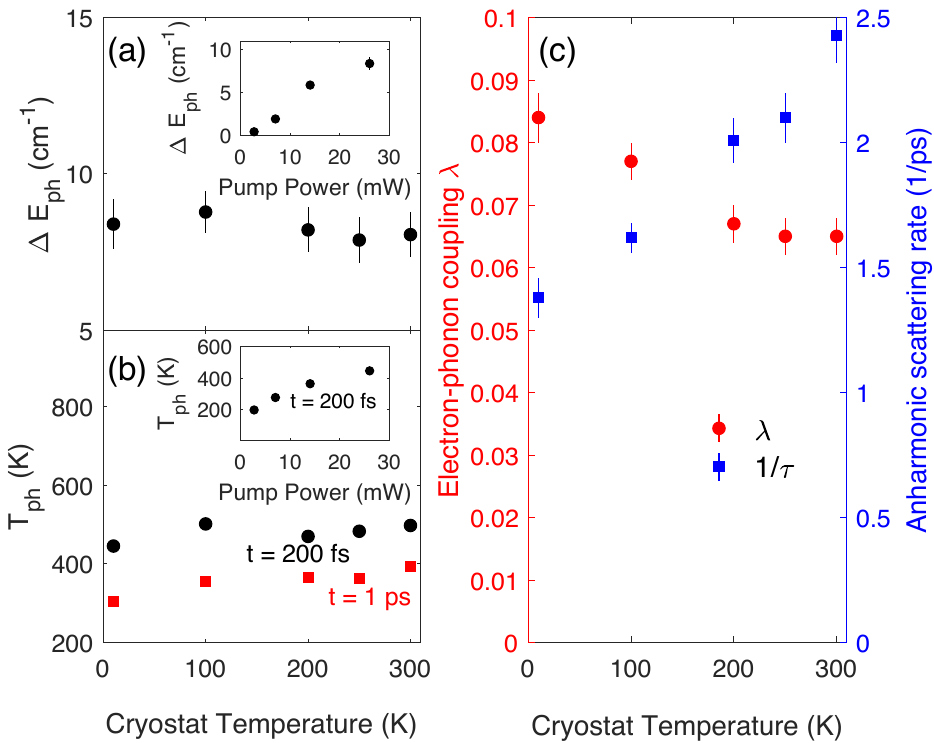} 
\caption{Dependence on pump power, P, and T$_{cryo}$. (a) Phonon hardening at 200 fs  (E$_{ph}$(200 fs) - E$_{ph}$(-1000 fs)) as a function of T$_{cryo}$ at P =  27 mW. Inset: Phonon hardening at T$_{cryo}$ = 10 K as a function of P. (b) Phonon temperature at 200 fs and 1 ps as a function of T$_{cryo}$ at P = 27 mW (fluence of 1.4 mJ/cm$^2$). Note that T$_{el}$ = T$_{ph}$ at 1 ps. Inset: T$_{ph}$ at 200 fs at T$_{cryo}$ = 10 K as a function of P. (b) Electron-phonon coupling constant $\lambda$ together with the anharmonic scattering rate 1/$\tau$.  See Supplementary Notes 2 and 3 for details on how $\tau$ and $\lambda$ were calculated \cite{Supp}}.
\label{PhononRenorm}
\end{figure} 

We interpret our results in terms of the two-temperature model where  $f$ is a fraction of phonons with strong electron-phonon coupling, $\lambda$. Photoexcited electrons first give off energy to the hot phonons  \cite{PhysRevLett.111.027403, PhysRevLett.95.187403, PhysRevLett.99.197001}, which in turn decay into other, cold phonons with the lifetime $\tau$ until all phonons and electrons thermalize at $T_{eq}$. The system then slowly equilibrates with the heat bath via propagating (mostly acoustic) phonons.  

Our experiments provide detailed information on each step of this process.

The electrons are pumped by a Gaussian pulse $P(t)$ with energy density $E_{pulse} = c_{tot}\Delta T$, where $\Delta T=T_{eq}-T_{cryo}$ and duration of 170 fs. The combined specific heat of electrons, hot phonons, and cold phonons, $c_{tot}$, depends on temperature, but we approximate it as the average of the values at $T_{cryo}$ and $T_{eq}$ taken from Ref. \cite{PhysRevLett.71.1740}. If $\Omega_0$ is a typical phonon energy, then the electronic temperature, $T_{el}$, the hot phonon temperature, $T_{h}$, and the cold phonon temperature, $T_{c}$, obey the rate equations
\begin{align}
\frac{\partial T_{el}}{\partial t} = -\frac{3\lambda\Omega_0^3}{\hbar\pi k_B^2}\frac{n(T_{el}) - n(T_{h})}{T_{el}} 
+ \frac{P(t)}{c_{el}(T_{el})},
\label{eltemp}
\end{align}
\begin{equation}
\frac{\partial T_{h}}{\partial t} = \frac{c_{el}(T_{el})}{c_{h}(T_{h})}\frac{3\lambda\Omega_0^3}{\hbar\pi k_B^2}\frac{n(T_{el}) - n(T_{h})}{T_{el}} - \frac{T_{h} - T_{c}}{\tau},
\label{hottemp}
\end{equation}
\begin{equation}
\frac{\partial T_{c}}{\partial t} = \frac{c_{h}(T_{h})}{c_{c}(T_{c})}\frac{T_{h} - T_{c}}{\tau},
\label{coldtemp}
\end{equation}
where $n(T) = \left(\textrm{e}^{\Omega_0/k_BT}-1\right)^{-1}$ and the specific heats are $c_{el} = \gamma T_{el}$, $c_h = 3 N_{atm} f \Omega_0 \left(\partial n_{h}/\partial T_{h}\right)$, and $c_c = 3 N_{atm} (1-f) \Omega_0 \left(\partial n_{c}/\partial T_{c}\right)$. Here, $N_{atm}$ is the number of atoms per formula unit ( = 12.9 for our sample) and the value of $\gamma$ was taken from measurements in Ref. \cite{PhysRevLett.71.1740}.  Equations \ref{eltemp}-\ref{coldtemp} correspond to equations 1-3 of Ref. \cite{PhysRevLett.99.197001} written down in terms of $\Delta T$, which we measured with high precision. 




We solved these coupled differential equations using the Euler method with 1 fs steps. Fig. \ref{temperatureandenergy}a presents a fit to this model with $\Omega_0$ = 60 meV and $\tau$ = 478 $\pm$ 25 fs (see Supplementary Note 3 for fits with different values of $\Omega_0$ and $\tau$ \cite{Supp}). $f$ = 0.14 and $\lambda$ = 0.065 were free parameters.  
Electrons initially heat up to 1000 K similarly to the previous result on BSCCO  \cite{PhysRevLett.99.197001}, and then cool quickly to thermalize with hot phonons at 300 fs. Thermalized hot electrons and hot phonons have a larger heat capacity compared with the heat capacity of electrons alone so they equilibrate with cold phonons with the much slower 478 fs time constant at 250 K. This lifetime increases with reduced temperature reaching 700 fs at 10 K (see Supplementary Note 2 and Fig. \ref{PhononRenorm}c in \cite{Supp}). It is much smaller than in graphite \cite{yang_novel_2017, PhysRevLett.54.2151, PhysRevB.80.121403, PhysRevLett.100.225503, kang2008optical, PhysRevB.81.165405, wu2012hot}. The electron-phonon coupling strength is close to $\lambda$ = 0.07 at all temperatures, which is smaller than the the LDA value \cite{Bohnen_2003}. 

Thermalization on subpicosecond timescales is controversial. Deviations from the Fermi-Dirac distributions of the electrons have been reported at pump-probe delay times of a few hundred fs in simple metals \cite{PhysRevLett.68.2834,PhysRevLett.68.2834,PhysRevB.102.214305}. However, the ARPES measurements of Ref. 40 showed that in BSCCO, thermalization occurs already at 100fs. It also showed that nonthermal distribution at shorter times is characterized by a small (only a few percent) deviation from the Fermi-Dirac distribution. Assuming that YBCO behaves similarly, for our purposes electrons can be treated as internally thermalized \cite{PhysRevLett.68.2834}. 

The assumption of internally thermalized hot phonons is valid only if they have similar electron-phonon coupling strength. For example, the 2T model breaks down in graphite where one hot phonon had a significantly larger electron-phonon coupling than another \cite{yang_novel_2017}. According to Ref. \cite{Johnston2010} as well as prior experimental work, the breathing phonons and the buckling phonons have a significantly stronger electron-phonon coupling than the apical oxygen mode \cite{reznik_phonon_2012,PhysRevLett.107.177004}. We found that adding phonons that have a much stronger electron-phonon coupling than the apical mode to the model can increase $\lambda$ by nearly an order of magnitude, thus $\lambda$ obtained from fits to the apical oxygen phonon occupation data alone is not accurate. Time-dependent electronic temperature, however, was only weakly model-dependent, which allowed us to make a quantitative comparison of the phonon energy vs electronic temperature to theory as discussed below.

We observe a profound shift of the phonon peak position to a higher energy (hardening) at small time delays (Fig. \ref{temperatureandenergy}b) with the maximum at 200 fs. It is independent of T$_{cryo}$ but decreases with reduced pump power (Fig. \ref{PhononRenorm}a). These changes in the phonon energy do not correlate with either phonon occupation or with the electric field of the pump, which should follow the cross-correlation curve in Fig. \ref{temperatureandenergy}c. Instead phonon hardening closely follows the electronic temperature. A similar effect in graphite was associated with electronic temperatures of over 1000 K \cite{PhysRevB.80.121403,Pogna2016,PhysRevLett.124.037401}. 

The phonon self-energy $\Pi(q,\Omega)$ is proportional to electronic polarizability, which depends on the electronic temperature \cite{Lamago_Cr,Abdurazakov_2018,Coslovich2013}. To make this quantitative, we model the system as a band of electrons interacting with the three strongly coupled phonons: the $A_{1g}$, $B_{1g}$ and breathing modes, which are kept at the cryostat temperature. Electronic quasiparticle excitations renormalize the apical phonon frequency. The increase in the quasiparticle temperature diminishes the softening of the apical phonon (lines in Fig.~\ref{schematics}b), leading to an increase in the phonon frequency as a function of the effective electronic temperature \cite{Abdurazakov_2018} (also see Supplementary note 4 in \cite{Supp} and references \cite{Johnston2010,Norman2007linear} therein for the details ).  The initial decrease of the theory curve at T$_{cryo}$=10K is a result of the broadening of the electronic bands due to their phonon-mediated self-energy. At weak interactions (and low T) broadening the bands increases the bubble (phonon self-energy), and thus softens the phonon as temperature increases.  This effect competes with the dominant hardening effect due to smearing of the Fermi surface described above. The calculation picks it up due to relatively sharp bands when the lattice temperature is low in the absence of impurity scattering. It is not present at 300K because of the existing broadening by phonons at this temperature. Increasing the impurity scattering rate weakens the kink at T$_{cryo}$=10K as shown in Supplementary note 5 \cite{Supp}.

Note that simple heating of the sample leads to phonon softening, not hardening. This is primarily because of thermal expansion caused by increased population of acoustic phonons. However, acoustic phonons stay cold at short pump-probe time delays and lattice expansion does not occur.

We have demonstrated quantitative agreement with theory of the time-varying apical phonon
frequency as the electronic system loses its excess energy to the broader phonon bath. It
highlights that the dynamics of energy transfer are responsible for the temporal behavior of the electrons
and phonons, as well as the disparity between interactions in- and out of equilibrium.
Although in strongly correlated materials such as YBCO there are strong Coulomb processes and impurity
scattering that dominate the electronic spectra, when it comes to time domain these processes rapidly come
to an internal equilibrium and effectively shut off \cite{kemper2018general}. Our results provide
a new, phonon-centered, perspective on previous experiment
in Bi$_2$Sr$_2$CaCu$_2$O$_8$ with both time-resolved ARPES \cite{rameau2016energy} and ultrafast electron diffraction \cite{konstantinova2018nonequilibrium} that both observed a similar quantitative agreement.

Our work provides new insights into photoinduced superconductivity \cite{PhysRevB.94.224303,PhysRevB.89.184516}. Its signatures were recently reported in the optical spectra of underdoped YBCO up to time-delays of about 1ps when pumping with 790 nm near-IR pulses \cite{Liu_2020} as well as with pulses that resonated with IR-active apical oxygen phonons. Our experiments reproduced the former pumping condition and showed that optimally-doped and underdoped YBCO behave similarly (Fig. \ref{raw}b). We found that hot and cold phonons were out of thermal equilibrium, but electrons and hot phonons were at or near thermal equilibrium at time delays below 1ps. At these time-delays electronic temperatures were always well above room temperature. We plan to determine transient heating while pumping the IR-active phonons in future experiments.



%
%



\section{Acknowledgements}
We would like to thank A. Cavalleri for a critical reading of the manuscript and helpful suggestions. Experiments at the University of Colorado were supported by the NSF under Grant No. DMR-1709946 and laboratory upgrade that made these experiments possible by DARPA through the DRINQS program. The work at University of Wisconsin-Madison (thin film synthesis and structural and electrical characterizations) was supported by the US Department of Energy (DOE), Office of Science, Office of Basic Energy Sciences (BES), under award number DE-FG02-06ER46327. A.F.K. acknowledges support from the National Science Foundation under Grant No. DMR-1752713. The work at the University of Wisconsin-Parkside was supported by WiSys and UW System Applied Research Grant award number 102-4-812000-AAH1775.

\bibliography{apical_refs}

\end{document}